\documentclass[aip,rsi,amsmath,amssymb,reprint]{revtex4-1}

\usepackage{graphicx}

\begin{document}

\title{A Fourier transform Raman spectrometer with visible laser excitation}

\author{S. Dzsaber}
\affiliation{Budapest University of Technology and Economics, Faculty of Physics, Budafoki \'{u}t 8., Budapest, H-1111, Hungary}
\affiliation{University of Vienna, Faculty of Physics, Strudlhofgasse 4., Vienna, A-1090, Austria}

\author{M. Negyedi}
\affiliation{Budapest University of Technology and Economics, Faculty of Physics, Budafoki \'{u}t 8., Budapest, H-1111, Hungary}

\author{B. Bern\'{a}th}
\affiliation{Budapest University of Technology and Economics, Faculty of Physics, Budafoki \'{u}t 8., Budapest, H-1111, Hungary}

\author{B. Gy\"{u}re}
\affiliation{Budapest University of Technology and Economics, Faculty of Physics, Budafoki \'{u}t 8., Budapest, H-1111, Hungary}

\author{T. Feh\'{e}r}
\affiliation{Budapest University of Technology and Economics, Faculty of Physics, Budafoki \'{u}t 8., Budapest, H-1111, Hungary}

\author{C. Kramberger}
\affiliation{University of Vienna, Faculty of Physics, Strudlhofgasse 4., Vienna, A-1090, Austria}

\author{T. Pichler}
\affiliation{University of Vienna, Faculty of Physics, Strudlhofgasse 4., Vienna, A-1090, Austria}

\author{F. Simon}
\email[Corresponding author: ]{ferenc.simon@univie.ac.at}
\affiliation{Budapest University of Technology and Economics, Faculty of Physics, Budafoki \'{u}t 8., Budapest, H-1111, Hungary}
\affiliation{University of Vienna, Faculty of Physics, Strudlhofgasse 4., Vienna, A-1090, Austria}

\date{\today}

\begin{abstract}
We present the development and performance of a Fourier transformation (FT)
based Raman spectrometer working with visible laser (532 nm) excitation. It is generally thought that FT-Raman spectrometers are not viable in the visible range where shot-noise
limits the detector performance and therein they are outperformed by
grating based, dispersive ones. We show that contrary to this
common belief, the recent advances of high-performance interference filters makes the
FT-Raman design a valid alternative to dispersive Raman spectrometers
for samples which do not luminesce. We critically compare the performance of our
spectrometer to two dispersive ones: a home-built single channel and a state-of-the-art CCD based instruments. We demonstrate a similar or even better sensitivity than the CCD based dispersive spectrometer particularly when the laser power density is considered. The instrument possesses all the known advantages of the FT principle of spectral accuracy, high throughput, and economic design. We also discuss the general considerations which helps the community reassess the utility
of the different Raman spectrometer designs.
\end{abstract}

\maketitle

\section{Introduction}

Raman spectroscopy \cite{ferraro2003introductory} is an important tool in various fields of science
from biology until physics or minearology. Raman spectroscopy is an
inelastic light scattering method, where the energy transfer to and
from the system under investigation is characteristic for electronic, optical, vibrational, or even magnetic properties \cite{KuzmanyBook}.\\

The availability of coherent light sources, i.e. lasers, lead to the proliferation of Raman spectroscopy and at present commercial instruments are available from several producers. Raman spectrometers are classified into dispersive and the Fourier transformation based ones depending on the way the spectrum of the scattered light is analyzed: the dispersion of optical gratings resolve light according to the wavelength, whereas the Fourier transformed signal from the interfering light in a Michelson-type optical interferometer is obtained for the latter. \\

Historically, dispersive spectrometers were developed first with single-channel photomultiplier (PMT) detectors \cite{smith2005modern}, where scattered photons for a single wavelength are measured at once. The development of charge coupled device (CCD) based multichannel detectors for visible light operation substantially improved the sensitivity of dispersive spectrometers as therein many wavelengths are measured simultaneously. It is now generally accepted that CCD based dispersive spectrometers are the best choice for visible Raman spectroscopy.

In principle, the Fourier transform (FT) based spectrometer (which was originally developed for infrared, IR, spectroscopy \cite{smith2011fundamentals}) would be a viable alternative for single-channel spectrometers or even for the CCD based ones. FT spectrometers measure all wavelengths \emph{simultaneously} which is known as the \textit{multiplex } or \textit{Fellgett's advantage} and it results in improved sensitivity compared to the single channel spectrometers \cite{Fellgett}. We denote the multiplex advantage by 'MA' for of the different spectrometer designs with respect to the single-channel dispersive spectrometers. Thus $\text{MA}(\text{CCD)}=\sqrt{N_{\text{pix}})}$ and $\text{MA}(\text{FT-IR})=\sqrt{N_{\text{ch}})}$ are the multiplex advantages for the CCD based visible Raman and the FT principle based IR spectrometers with $N_{\text{pix}}$ and $N_{\text{ch}}$ being the number of CCD pixels and measured time bins of the two kinds of spectrometers, respectively.

FT-Raman spectrometers were developed as late as in 1986 due to the reasons discussed herein \cite{ChaseJACS1986,ChaseHirschfeld}. It is known that shot noise dominates the noise of visible detectors, i.e. for an FT system the noise power is spread out over all wavelengths at once thus the multiplex advantage is lost or such a spectrometer would perform even worse than a dispersive one if the source has fluctuation (or flicker) noise \cite{Chase2003}. It was however recognized by Chase and Hirschfeld \cite{ChaseHirschfeld} that FT-Raman does have an advantage for near-IR (NIR) Raman spectroscopy where noise is due to that of the detector and thus the multiplex advantage is present \cite{Hirschfeld1}. In general, a NIR Raman spectrometer suffers from the low Raman cross section due to the $1/\lambda^4$ rule \cite{KuzmanyBook} but it is balanced by the advantages of the FT-Raman principle and the absence of luminescence for NIR excitation.

Nevertheless, the overall judgement that "FT-Raman spectrometers are not viable for visible excitation" stuck and it still hinders development in this direction \cite{VisFTRaman_negative,EverallHoward,BrenanHunter,McCreery1996,McCreery1997}. E.g. Savoie and co-workers \cite{VisFTRaman_negative} found that a visible FT-Raman spectrometer operates with a factor 5 lower sensitivity than a single-channel PMT based dispersive one.

To understand this generic assessment about visible FT-Raman spectrometers, we consider the shot-noise for a PMT: it is due to the total flux of photons, $N$. The photon flux is converted to a cathode current, $I_{\text{c}}$ which consists of three contributions from dark current, the desired signal, and that of unwanted light. These contributions add incoherently and their noise contributions appear uniformly over all detected wavelengths after Fourier transformation. The dark current can be usually diminished to a low level by detector cooling. If unwanted light is present, e.g. due to Rayleigh scattering or luminescence, even at a wavelength which does not overlap spectrally with the desired signal, its noise will overwhelm the spectrum. Therefore the key factors to perform visible FT-Raman spectroscopy are to i) work on samples without luminescence and ii) eliminate all unwanted light: environment background, Rayleigh radiation, and the light of the He-Ne acquisition laser of FT-Raman spectrometers.

Extending the FT-Raman spectrometer operation to the visible range would be of great advantage due to the $1/\lambda^{4}$ wavelength dependence of the Raman process \cite{KuzmanyBook}. In addition, the FT-Raman operation has several other advantages compared to the dispersive technique \cite{Chase2007}: i) wavelength is calibrated with respect to a coherent laser (known as the Connes' advantage), which eliminates the tedious calibration process of dispersive spectrometers, ii) FT spectrometers do not require an input slit, i.e. all photons are measured (known as the throughput or Jacquinot's advantage), iii) the exciting light does not need to be focussed on the sample, which allows to increase the incident laser power, iv) resolution in FT spectrometers is readily modified whereas only predefined, discrete resolution values are available for dispersive ones at the cost of grating change, v) high resolution FT-Raman spectrometers have a smaller footprint than dispersive ones, vi) use of a PMT is more economical than that of a liquid nitrogen cooled CCD detector.\\

Herein, we present the development of an FT-Raman spectrometer which operates with a visible laser and its detector is a PMT at room temperature. We show that while it has all the usual FT advantages, it does not suffer substantially from the shot-noise problem. Our spectrometer exploits the recent developments in the field of high quality interference filters which allow the suppression of unwanted light by up to 7 orders of magnitude (OD7), which leads to reduced shot-noise. We compare the performance of the spectrometer to a commercial, state-of-the-art dispersive spectrometer using sulfur powder and we show that the two have similar sensitivity and when the laser power density on the sample is considered, the FT-Raman spectrometer outperforms the dispersive one by orders of magnitude. We revisit the general considerations which lead to the conclusion that shot-noise prevents successful operation of FT-Raman spectrometers for the visible range and we show under what circumstances this problem is not significant. We also discuss the overall design of the several Raman spectrometer types on general grounds in order to help reassess their utility.

\section{The spectrometer and its performance}
\subsection{The spectrometer setup}

Following the overall principles in instrumentation, the setup was motivated to maximize the signal and minimize the noise of our spectrometer. The earlier is achieved by using i) an aberration free, high-speed (i.e. small $f/\#$) light-collection objective and ii) optical elements (mirrors and beamsplitter) and detector which is optimized for visible light. Since shot-noise limits the performance of our spectrometer, unwanted light sources has to be eliminated. This is achieved by using three high performance interference filters, a pinhole, and a darkened environment.

\begin{figure}[htp]
\begin{center}
\includegraphics[width=0.5\textwidth]{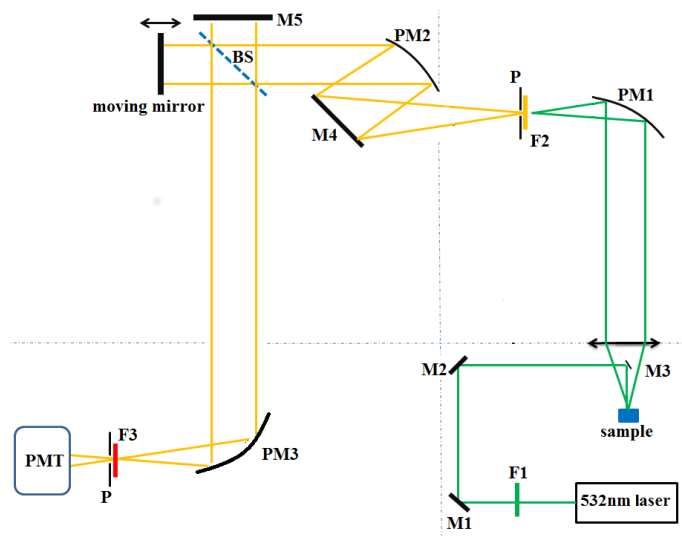}
\caption{Schematics of the visible laser based FT-Raman spectrometer.}
\label{Fig:block}
\end{center}
\end{figure}

The spectrometer setup is shown in Fig.~\ref{Fig:block}. It is based on a commercial infrared Fourier-transform (FT-IR) and Fourier-Raman (FT-Raman) spectrometer (\emph{Bruker IFS66v with FRA 106 Raman module}). The FT-Raman spectrometer was optimized for excitation with a 1064 nm Nd:Yag laser, i.e. for the NIR range. Several optical elements including the exciting laser, collecting objective, guiding optics, beam splitter, and detector were replaced with ones suitable for operation in the visible optical range. The spectrometer is divided into four functional parts: sample compartment, light collecting compartment, the interferometer, and the detector compartment.

A frequency doubled Nd:YAG solid state laser at $532$ nm is used (\emph{Optotronics Inc. VA-I-100-532 100 mW}) for excitation with a beam diameter of 1.2 mm. A $650$ nm short pass filter, F1 (\emph{Thorlabs FES0650}), removes the $812$ nm radiation output from the laser. The laser beam is guided to the sample with two dielectric mirrors (M1 and M2) and a small (3x3 mm) right angle prism M3, (\emph{Edmund Optics \#47-921}). This is the usual setup for the so-called macro-Raman configuration, i.e. when the excitation and the scattered Raman light are separated (or duplexed) according to the different beam sizes. We note that the light is not focused on the sample in this configuration. This is possible for FT-Raman spectrometers, where the light does not need to come from a well focused source as their is no input slit in contrast to dispersive spectrometers. Resolution in FT spectrometers depends on the mirror movement of the interferometer rather than on the slit size, i.e. more light can be collected and also the laser power density on the sample is substantially smaller. This effect is known as \textit{throughput or Jacquinot's advantage} of FT-Raman spectrometers and is discussed in depth below.

The light is scattered from the sample in a $180^{\circ}$ geometry and is collected with a state-of-the-art, eight element double-Gaussian objective lens (50 mm $f/0.95$, \emph{Navitar DO-5095} of eight optical elements). The objective produces a 1.8'' diameter beam which is focused by a 2'' $90^{\circ}$ off-axis parabolic mirror, PM1 (\emph{Edmund optics, NT63-186, Aluminium coated}) on a pinhole, P, of 2 mm diameter. There is a 1'' diameter $532$ nm long-pass filter (LPF), F2 (\emph{Semrock  LP03-532RE-25} before the pinhole. The LPF has optical density 7 (OD7) rejection for the stop-band and $>93\,\%$ transmission for the pass-band. The transition edge of the filter can be fine tuned by rotating around a vertical axis \cite{FabiRSI} as the edge blue shifts when rotated away from normal incidence. The LPF efficiently filters out the undesired quasi-elastic (near 532 nm) radiation, which is commonly referred to as \textit{Rayleigh radiation} and it allows only the Stokes Raman radiation to pass. The light is incident on the interference filter with a maximum angle of $\theta=8^{\circ}$ which broadens the $tr=93\,\text{cm}^{-1}$ transition range of the LPF (where transmission changes between OD6 and OD0) according to Ref. \cite{InterferenceFilter}:

\begin{gather}
tr'=tr+ \lambda_0 \left(1-\sqrt{1-\frac{sin^2 \theta}{n^{*2}}}\right),
\label{Eq:transition_range}
\end{gather}

\noindent where $n^* \sim 1.5$ is the index of refraction of the filter material and $\lambda_0=18797\,\text{cm}^{-1}$. Eq.\eqref{Eq:transition_range}. yields a somewhat larger transition range of $tr'=133\,\text{cm}^{-1}$ of the LPF, which does not affect its performance. The use of this moderately focused geometry eliminates the need for an LPF with 2'' diameter. We found that the order of the LPF and the pinhole is important: light reflected from the front side of the LPF can reflect toward the spectrometer for a reversed order, which deteriorates the performance. We found that the pinhole reduces the intensity level of the Rayleigh light by a factor of 3 without affecting the Raman signal.

The interferometer compartment was unchanged with respect to the commercial setup. Therein the Raman light is collimated and guided to the interferometer with a planar, M4, and a parabolic mirror, PM2. A Michelson-type interferometer produces the interferogram with the standing mirror, M5, a moving mirror, and a quartz beam splitter, BS (\emph{Bruker T502/1}) optimized for 470-833 nm). We used 5 kHz scanning speed of the interferometer but other values between 500 Hz and 10 kHz are possible with the IFS66. The intensity modulated beam produced by the interferometer enters the detector compartment where it is focused by the parabolic mirror PM3 on a $633\,\text{nm}$ short pass filter (SPF), F3 (\emph{Semrock SP01-633RU}), to eliminate the $632.832\,\text{nm}$ line of the He-Ne acquisition laser. The optical bandwidth (OBW) of the spectrometer is thus given by the two filters (532 nm LPF and 633 nm SPF) and is $532-633\,\text{nm}$. This OBW corresponds to a maximum of $3000\,\text{cm}^{-1}$ Stokes Raman shift with respect to the 532 nm excitation, which is sufficient for most Raman studies. The range could be extended towards longer wavelengths by using a stop-band filter instead of the 633 nm short pass filter. Alternatively, our spectrometer could be readily modified to detect anti-Stokes Raman scattering by replacing the 532 nm long-pass filter by a short pass filter.

A photo multiplier tube, PMT, (\emph{Hamamatsu R955}, 160-900 nm, Q.E.$\approx 10\%$) followed by an I/V converter (\emph{Hamamatsu C7319}) is used to detect the interferogram. The PMT cathode-anode voltage can be set up to 1500 V and the corresponding PMT gain is obtained from its datasheet. The I/V converter can be set for either 20 or 200 kHz bandwidth (BW) and $10^5$-$10^7$ gain. The IFS66 instrument internally sets a BW that is near the scanning speed of the interferometer, which is optimal for this measurement. The latter information was obtained by comparing the noise in the signal digitized by the IFS66 to that obtained using an external analog-to-digital converter. Noise of a PMT is known to be due to shot noise and the signal-to-noise ratio for a given cathode current, $I_{\text{c}}$, is [Ref. \onlinecite{RobbenPMTNoise}]:

\begin{gather}
S/N(I_{\text{c}})=\sqrt{\frac{I_{\text{c}}}{2e \text{BW}}},
\label{Eq:Shot_noise}
\end{gather}

\noindent where $e$ is the elementary charge. We note that Eq. \eqref{Eq:Shot_noise}. is valid irrespective of the origin of the cathode current as it can be due to light or due to dark current (thermal fluctuations or cosmic radiation).

\subsection{Performance of the VIS FT-Raman spectrometer}

\begin{figure}[htp]
\begin{center}
\includegraphics[width=0.5\textwidth]{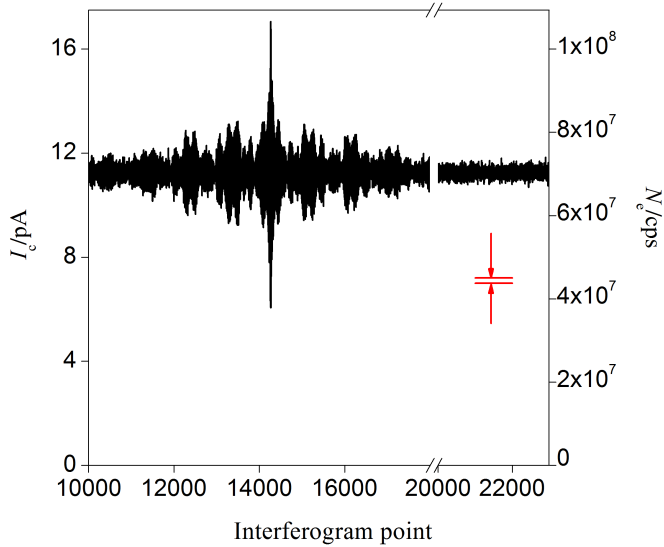}
\caption{Interferogram of sulfur powder given in units of the PMT cathode current. The corresponding number of cathode electrons is also given. The noise ($\sigma=0.2\,\text{pA}$) of the current is obtained from the part of the interferogram where the coherent light has decayed and its magnitude is shown between arrows (offset for clarity). The fringe visibility is about $0.5$, indicating the presence of unmodulated light on the detector.}
\label{Fig:interferogram}
\end{center}
\end{figure}

To characterize the performance of the visible FT-Raman spectrometer, we carried out Raman measurements on sulfur powder as it has a relatively strong Raman response and is often used as a benchmark sample. The corresponding Raman interferogram is shown in  Fig.~\ref{Fig:interferogram}. It was recorded with $10$ mW incident laser power, 5 kHz scanner velocity, a single scan, and $4\,\text{cm}^{-1}$ spectral resolution. Acquiring the interferogram takes $0.9$ seconds under these conditions.
The cathode current data, shown in Fig.~\ref{Fig:interferogram}., was obtained from the I/V converter voltage output and the value of the known PMT gain. Its average value is $I_{\text{c}}=11.2\,\text{pA}$ and has a Gaussian noise spectrum with $\sigma=0.2\,\text{pA}$ as deduced from the parts of the interferogram without interference fringes.

The quality of the interferometer is usually characterized by the so called fringe visibility of the interferogram, given by:

\begin{equation}
\nu = \frac{I_{\text{max}}-I_{\text{min}}}{I_{\text{max}}+I_{\text{min}}},
\label{Eq:fringevisibility}
\end{equation}

\noindent where $I_{\text{max}}$ and $I_{\text{min}}$ are the maximum and minimum values of the interferogram, respectively. For an ideal case $\nu=1$, i.e. the total light which enters the detector comes from coherent radiation which interfere in the two arms of the interferometer. We found that $\nu \sim 0.5$ in our case which is a typical value for visible FT-Raman spectroscopy \cite{VisFTRaman_negative} and it indicates that about half of the total intensity which reaches the PMT comes from the modulated light. We found that the unmodulated portion of the incoming light scales with the laser power, i.e. it does not come from a background. The fringe visibility can be as high as 0.8 for the infrared operation; the lower value in the shorter wavelength visible is related to the sensitivity of the interferometric principle to surface imperfections and misalignments.

As discussed above, the cathode current defines the noise and the resulting signal-to-noise ratio according to Eq.\eqref{Eq:Shot_noise}. This gives S/N=86 with $I_{\text{c}}=11.2\,\text{pA}$ and BW=5 kHz. This is in good agreement with the experimental value of S/N=56 as obtained from Fig. \ref{Fig:interferogram}. The agreement proves that the noise is indeed due to shot-noise in the interferogram and that additional noise sources, e.g. laser flicker noise, are absent. We note that the dark current of the PMT (at room temperature) is 5 fA, i.e. three orders of magnitude smaller than the current due to light, thus its contribution to noise is negligible. Cooling of the PMT might be necessary to lessen the dark current when a lower light flux is to be detected.

\begin{figure}[htp]
\begin{center}
\includegraphics[width=0.5\textwidth]{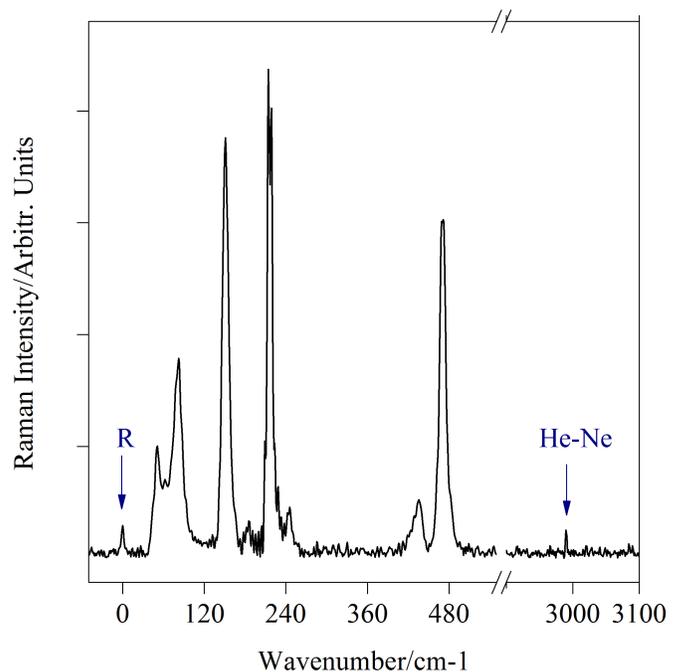}
\caption{\begin{small}
Raman spectrum of sulfur powder recorded with the  visible FT-Raman spectrometer. The spectrum was obtained from the interferogram shown in Fig. \ref{Fig:interferogram}. Arrows for "R" and "He-Ne" show the residual modes from the Rayleigh scattering and the He-Ne acquisition laser.
\end{small}  }
\label{Fig:spectrumFT}
\end{center}
\end{figure}

The spectrum shown in Fig.~\ref{Fig:spectrumFT}. is obtained from the interferogram in Fig. \ref{Fig:interferogram}. by a Fourier transformation. The characteristic peaks of sulfur \cite{SulphurRamanJPhysChem} at 86, 147, 188, 215, 247, 434, and 472 $\text{cm}^{-1}$ are readily recognized. We also observe small, residual peaks due to the Rayleigh radiation and the He-Ne acquisition laser. Were these wavelengths not effectively attenuated in our measurement, they would overwhelm the spectrum and would give rise to a giant shot-noise.

We consider the ratio of the largest sulfur peak amplitude (at $218\,\text{cm}^{-1}$) to the noise which gives a signal-to-noise-ratio of the spectrum $S/N=155$. The noise is obtained from parts of the spectrum without peaks, i.e. $750\,\text{cm}^{-1}$ - $1500\,\text{cm}^{-1}$. The increase of $S/N$ compared to that of the interferogram ($S/N=56$) is a factor of three and it is due to the well known property of Fourier transformation that it improves $S/N$. Although related, it is not identical with the Fellgett or multiplex advantage.

\subsection{Comparison to dispersive spectrometers}

The performance of our VIS-FT spectrometer was compared to two dispersive Raman spectrometers: a home-built single-channel and a CCD based multi-channel spectrometer. The earlier was based on a 320 mm spectrograph equipped with a 2400 grooves/mm grating (\emph{Jobin  Yvon iHR320}) and the same $f/0.95$ objective and laser that was used for the visible FT-Raman spectrometer.
The latter was a state-of-the-art commercial Raman spectrometer (\emph{Horiba Jobin-Yvon: LabRAM-HR800}) ~\cite{FabiRSI} equipped with a $532$ nm laser and a 1024 pixel CCD photo-detector (\emph{Horiba Symphony II}) a 600 grooves/mm grating and 800 mm focal length. The collecting objective was a microscope (\emph{Olympus LMPlan 50x/0.50, inf./0/NN26.5}) with N.A.=0.5 that yields about $1.3 \times 1.3,\mu m^2$ spot size.
For a valid comparison, the same sulfur sample was measured for the same resolution, $4\,cm^{-1}$, optical bandwidth, $3000\,cm^{-1}$, and a laser power of $10\,\text{mW}$ as in case of the visible FT-Raman spectrometer. The role of the laser power density for the different systems is discussed further below. For the multi-channel dispersive spectrometer, the optical grating and CCD covers $\sim 1500\,\text{cm}^{-1}$ and it thus takes two measurements to cover the $3000\,\text{cm}^{-1}$ Raman spectrum.

\begin{figure}[htp]
\begin{center}
\includegraphics[width=0.5\textwidth]{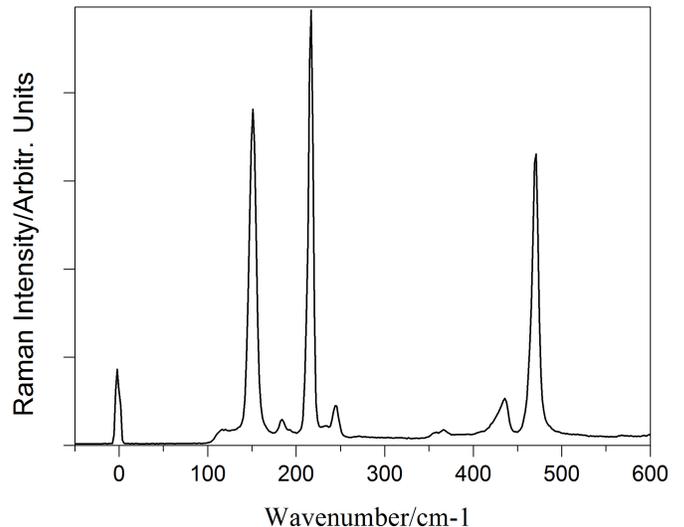}
\caption{Raman spectrum of sulfur powder excited at 532 nm, measured with the CCD based dispersive spetrometer. The experimental conditions are discussed in the text. Note the edge of the long-pass filter near $110\,\text{cm}^{-1}$.}
\label{Fig:Labram_spectrum}
\end{center}
\end{figure}

The Raman spectra of sulfur powder is shown in Fig. \ref{Fig:Labram_spectrum}. as obtained with the CCD based dispersive spectrometer (data obtained with the single-channel spectrometer is not shown). Again, we obtain the signal amplitude from the largest sulfur peak and the noise from the standard deviation of the signal, $\sigma$ for a spectral range where no Raman peaks are present. When normalized by the same measurement time as for the FT-Raman data, we obtain $S/N_{\text{single-channel}}=19$ and $S/N_{\text{CCD}}=815$ for the single-channel and CCD based dispersive, spectrometers, respectively. For the CCD based spectrometers, the measurement involved two rotated grating positions. These figures are to be compared with the $S/N_{\text{FT}}=155$ for the FT-Raman spectrometer. This clearly shows the superior performance of the FT-Raman spectrometer with respect to the single-channel dispersive one and hints at an inferior performance as compared to the CCD-based dispersive spectrometer. This comparison did not yet consider the role of the laser power density, which is substantially larger for dispersive spectrometers than for the FT principle based ones, and is discussed in detail below.

Another important difference between the two types of spectrometers is the distribution of the shot noise. While the apparent noise is the noise of the background for the dispersive spectrometer, we show herein that spectroscopically a larger noise is to be considered, which reduces the apparent advantage of the CCD based technique. The CCD spectrometer is also affected by shot noise, i.e. the noise increases for an observed peak as $\sigma=\sqrt{S}$, where $S$ is the signal in cps units and $\sigma$ is the standard deviation. The apparent noise of the background is usually much smaller and is typically due to dark current and is about $\sigma_{\text{D}} \approx 3-10\,\text{cps}/\sqrt{\text{Hz}}$ for the LabRam instrument. However, when the spectroscopic parameters of a Raman signal are determined by fitting (line position, intensity, and line width), the resulting errors are affected by the shot noise of the signal itself and not merely by that of the background. As a result, one deals with an \textit{effective noise} which is larger than that of the background alone.

We performed a numerical simulation for a hypothetic Raman peak of $S\,\text{[cps]}$ amplitude and spectroscopic width of $L$ pixels. We found that the resulting spectroscopic parameters have an error which would be obtained for a uniform total noise, $\sigma_{\text{t}}$ of:

\begin{gather}
\sigma_{\text{t}}=\sqrt{\sigma_{\text{D}}^2+\frac{S}{L}}.
\label{Eq:NoiseProp}
\end{gather}

Eq. \eqref{Eq:NoiseProp}. means that the spectroscopically relevant noise is larger than the apparent one. For sulfur, the typical $S=10,000\,\text{cps}$ was observed for $L \approx 10$ and we obtain $\sigma_{\text{t}}\approx 30$ which is a factor 3-10 larger than the apparent noise. We note that this kind of extra noise is not additive for the different peaks, i.e. it increases the spectroscopic uncertainty of each peaks independently. Nevertheless, the difference between the presence of an effective noise, which is larger than the apparent one, mean that the seemingly better performance of the CCD based spectrometer is somewhat reduced.

\section{Discussion of the Raman spectrometer design}
\subsection{The multiplex advantage}

The overall assessment on the magnitude of the multiplex advantage is summarized in Table \ref{tab:Comparison}. for the following spectrometer design: dispersive spectrometer with a CCD for the visible range (disp.-CCD/VIS), FT-Raman spectrometer for the NIR range (FT-Raman/NIR) and FT-Raman spectrometer for the visible range (FT-Raman/VIS). We compare the multiplex advantage to the corresponding single-channel spectrometers: dispersive single-channel for the visible (disp.-sch/VIS) and NIR (disp.-sch/NIR). The number of CCD pixels is $N_{\text{pix}}$, the number of time domain bins, $N_{\text{ch}}$ is obtained from the resolution of the FT-Raman method. The Nyquist frequency of the data sampling corresponds to $\nu\approx15803\,\text{cm}^{-1}$ of the He-Ne laser, thus a $4\,\text{cm}^{-1}$ resolution implies $N_{\text{ch}}\approx 4000$. We inserted a factor of 5 into the relevant values in Table \ref{tab:Comparison}. to reflect the larger quantum efficiency of CCD's ($\sim 50\%$) as compared to PMT's ($\sim 10\%$). $G$ is the number of measurement windows which are required to cover the desired spectral range with the dispersive spectrometer.

The multiplex advantage depends on the type of spectrum which is measured with the FT-Raman/VIS technique. It is 1 (i.e. no advantage) when the spectrum is broadband \cite{ChaseHirschfeld,Hirschfeld1} and has a maximum multiplex advantage of $\sqrt{N_{\text{ch}}}$ for a hypothetical Raman spectrum consisting of a single line whose line-width matches that of the instrument resolution. For a Raman spectrum consisting of $P_{\sharp}$ lines of equal amplitude and each having $w$ spectral line-width which is $L$ times larger than the spectral resolution, we obtain $\sqrt{\frac{N_{\text{ch}}}{P_{\sharp}L}}$ for the multiplex advantage. This formula recovers the extremal cases of the single peak ($P_{\sharp}=1$ and $L=1$) and the broadband spectrum ($P_{\sharp}=1$ $L=N_{\text{ch}}$). The multiplex advantage value can be similarly obtained for an arbitrary Raman spectrum.

In the case of our measurements with the disp.-CCD/VIS and FT-Raman/VIS spectrometers for sulfur, the relevant values are: $\sqrt{5\,N_{\text{pix}}/G}\approx 50$ ($N_{\text{pix}}=1024$, $G=2$) and $\sqrt{\frac{N_{\text{ch}}}{P_{\sharp} L}}\approx 14$ ($N_{\text{ch}}=4000$, $P_{\sharp} L\approx 20$), which have to be compared with the corresponding experimental values of 42 and 8, respectively. We find that the calculated and measured multiplex advantage values are in good agreement for both kinds of spectrometers. This also implies that our visible FT-Raman spectrometer properly functions, i.e. there is e.g. no substantial optical loss.

\subsection{General considerations of the Raman spectrometer design}

We discuss the alternatives of the spectrometer designs for the visible range including CCD based macro and micro Raman spectrometers and the present development of a visible FT-Raman spectrometer. The difference between micro and macro Raman spectrometers is that for the earlier, the sample is irradiated with a diffraction limited spot size (about 0.5-1 microns diameter, the Airy disk), whereas for the latter the irradiation size is about 10-30 times larger. For a macro spectrometer, the objective lens with diameter $d$ is illuminated with a smaller beam diameter of $d_{\text{b}}$ and the resulting spot size is $d/d_{\text{b}}$ times larger than the Airy disk diameter. It is known that micro Raman spectrometers cause sample heating or damage for laser powers above about $10\,\text{mW}$.

Focusing the exciting laser beam is required for dispersive spectrometers as therein the resolution is defined by the width of the incoming slit (typically $100\,\mu\text{m}$): were the Raman light not collected from a focused spot, the transmission of the slit would be severely limited. We note that micro Raman spectrometers have a larger throughput compared to a macro one: the collection optics magnifies the image of the irradiated spot by 5-10 times (known as $f/\#$ matching \cite{smith2005modern}) i.e. the entrance slit becomes an aperture stop for macro Raman spectrometers. This higher throughput and high lateral resolution explain why the micro Raman spectrometers became more favored recently.

The FT-Raman principle, however does not require focusing of the irradiation as there is no input slit and the spectrometer resolution depends solely on the travel of the moving mirror of the interferometer. Using a non-focused irradiation of the sample has some limitation though as its size gives the angle of deviation from a perfectly parallel beam\cite{SmithModernOpticalEng} (also known as field angle):

\begin{gather}
\alpha=\arctan \frac{h}{2\cdot\text{EFL}}
\end{gather}

\noindent where $h$ is the height of the irradiated spot and EFL is the effective focal length of the objective. For our case of  EFL=50 mm and beam diameter of 1.2 mm, we obtain $\alpha=0.57^{\circ}$ which results in a beam expansion of 10 mm for a 1 m optical length. As a result, the unfocused beam does not result in a substantial intensity loss in our case, in agreement with the above discussed sensitivity of our spectrometer. We note that a convenient property of the Michelson-type interferometer is its insensitivity for the direction of the incoming light.

An attractive feature of the ability to use an unfocused laser beam is the substantially lower laser power density: it amounts to $10^{8}$ when a 1 mm diameter beam is compared to a $1\,\mu\text{m}$ spot size. This enormous room to increase the incident laser power on the sample compensates for the somewhat lower sensitivity of the herein presented visible FT-Raman spectrometer as compared to the CCD based dispersive one. We believe that a value of a few 100 mW incident power might be a reasonable compromise when other effects, such as e.g. heat transfer from the sample, is not limiting while the signal is as large for the FT-Raman spectrometer than for a CCD based dispersive one.

This concludes our comparison of the different spectrometer considerations: when high lateral resolution is required, a CCD based dispersive Raman spectrometer equipped with a microscope is the best choice. However, when larger samples are available, the FT-Raman principle provides a better alternative if ample exciting laser power is available and the spectrum contains well defined Raman peaks and is free from a broadband radiation.

\section{Conclusions}

In conclusion, we presented the construction and characterization of a Fourier transformation based spectrometer working with visible laser excitation. We have identified under what circumstances such a spectrometer is a viable alternative for CCD based multichannel Raman spectrometers which operate with gratings. This contradicts the common perception that FT-Raman spectrometer can by no means function as well as a dispersive multi-channel instrument. The performance of the present spectrometer is proven with measurements on sulfur powder. Given the other known advantages of the FT principle of e.g. accuracy, smaller footprint, and sensitivity, the present instrument may promote the development of a new class of Raman spectrometers.

\begin{acknowledgments}
Work supported by the European Research Council ERC-259374-Sylo, Austrian Science Funds (FWF), the APART grant Nr. 11456 of the Austrian Academy of Sciences, and Hungarian Science Funds (OTKA) K101778 grants. The authors are indebted to I. K\'{e}zsm\'{a}rki for stimulating discussions and for lending the photomultiplier and the quartz beamsplitter. R. Hackl is acknowledged for enlightening discussions about the construction of Raman spectrometers.
\end{acknowledgments}

\section{References}
\bibliographystyle{unsrt}
\bibliography{reference}
\pagebreak

\begin{table*}[htp]
\begin{center}
    \begin{tabular*}{0.5\textwidth}{@{\extracolsep{\fill} }l    l    l}
    \hline \hline
     & \begin{tabular}{rr}multiplex\\ advantage\end{tabular} & \begin{tabular}{rr}reference single\\ channel method\end{tabular}\\ \hline
    FT-Raman/NIR & $\sqrt{N_{\text{ch}}}$ & disp.-sch/NIR\\
    disp.-CCD/VIS & $\sqrt{5\,N_{\text{pix}}/G}$ & disp.-sch/VIS\\
    FT-Raman/VIS &  & \\
        \begin{tabular}{ccc}single peak\\broadband\\Raman-like\end{tabular}&\begin{tabular}{ccc}$\sqrt{N_{\text{ch}}}$\\$1$\\$\sqrt{\frac{N_{\text{ch}}}{P_{\sharp} L}}$\end{tabular} & $\Bigg\}$disp.-sch/VIS\\
    \hline \hline
    \end{tabular*}
    \caption{The multiplex advantage factor for the different techniques with respect to the corresponding single-channel detection method. We give values for three separate cases (single peak, broadband, and Raman-like spectra) for the FT-Raman/VIS spectrometer. The symbols, $N_{\text{pix}}$, $N_{\text{ch}}$, $G$, $P_{\sharp}$, $L$ and the spectrometer types are defined in the text.}
    \label{tab:Comparison}
\end{center}
\end{table*}
\pagebreak

\end{document}